\documentclass[11pt]{article}
 \usepackage[margin=1in,footskip=0.25in]{geometry}
 \usepackage{amssymb}
\usepackage[T1]{fontenc}
\usepackage[latin9]{inputenc}
\usepackage{array}
\usepackage{booktabs}
\usepackage{mathrsfs}
\usepackage{amsmath}
\usepackage[graphicx]{realboxes}
\usepackage{esint}
\usepackage{bm}
\usepackage{hyperref}
\usepackage{mathtools}
\usepackage{lscape}
\usepackage[figuresleft]{rotating}
\usepackage{setspace}
\usepackage[sort&compress, numbers]{natbib}
\usepackage[linesnumbered,ruled,vlined]{algorithm2e}
\usepackage{authblk}
\usepackage{lscape}
\usepackage{xcolor}
\begin{document}
\title{New travelling wave solutions of the Porous-Fisher model with a moving boundary}

\footnotesize
\author{Nabil~T.~Fadai$^*$, Matthew~J.~Simpson}
\affil{\footnotesize{School of Mathematical Sciences, Queensland University of Technology, Brisbane, Queensland 4001, Australia.
*Corresponding author email address: \texttt{nabil.fadai@qut.edu.au}
}} 
 \maketitle

\begin{abstract}
We examine travelling wave solutions of the Porous-Fisher model, $\partial_t u(x,t)= u(x,t)\left[1-u(x,t)\right] +  \partial_x \left[u(x,t) \partial_x u(x,t)\right]$, with a Stefan-like condition at the moving front, $x=L(t)$. Travelling wave solutions of this model have several novel characteristics.  These travelling wave solutions: (i) move with a speed that is slower than the more standard Porous-Fisher model, $c<1/\sqrt{2}$; (ii) never lead to population extinction; (iii) have compact support and a well-defined moving front, and (iv) the travelling wave profiles have an infinite slope at the moving front. Using asymptotic analysis in two distinct parameter regimes, $c \to 0^+$ and $c \to 1/\sqrt{2}\,^-$, we obtain closed-form mathematical expressions for the travelling wave shape and speed. These approximations compare well with numerical solutions of the full problem.
\end{abstract}


\section{Introduction}

\begin{figure}
\begin{center}
\includegraphics[width= \textwidth]{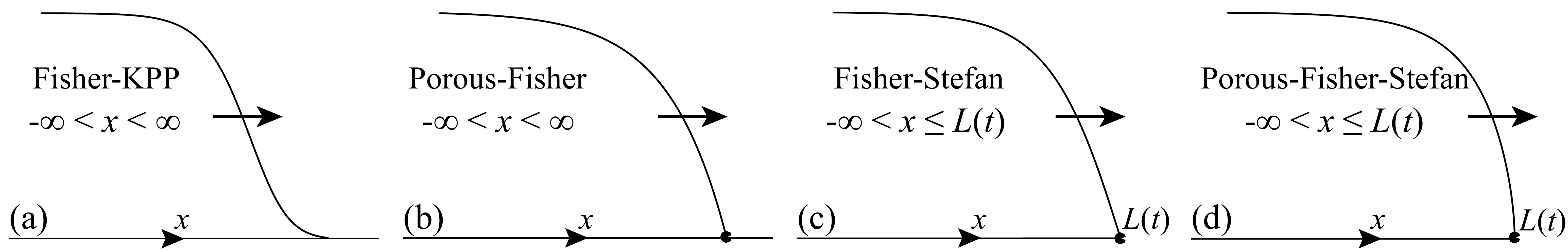}
\end{center}
\vspace{-0.2in} 
 \caption{Schematic representation of various monotone travelling waves. \label{fig:schematic}}
\end{figure}

Travelling waves arise in many fields, including ecology \cite{murray, fisher1937wave, holmes1994partial, bao2018free}, cell biology\cite{sherratt1990models, mccue2019hole, el2019revisiting, maini2004travelling, maini2004traveling, simpson2006looking, simpson2011models}, and industrial applications involving heat and mass transfer \cite{mcguinness2000modelling, dalwadi2017mathematical, fadai2018asymptotic, brosa2019extended}. Such processes are often modelled using reaction-diffusion equations and, depending on the choice of reaction and diffusion terms, three broad classes of monotone travelling waves are commonly reported (Figure \ref{fig:schematic}). The most commonly-observed travelling wave is a smooth front (Figure \ref{fig:schematic}a), whereby the concentration, $u(x,t)$, is a monotone decreasing function decaying to zero as $x\to\infty$. For example, travelling wave solutions of the Fisher-KPP model \cite{murray, fisher1937wave, el2019revisiting},
\begin{equation}
\partial_t u(x,t)= u(x,t)\left[1-u(x,t)\right] +  \partial_{xx} u(x,t), \qquad -\infty < x < \infty,
\end{equation}
are smooth.
 Unfortunately, smooth fronts do not have compact support, which makes defining the ``edge'' of the moving front ambiguous \cite{maini2004travelling, maini2004traveling, treloar2013sensitivity}. This feature of the Fisher-KPP model means that it can be hard to apply to practical problems, such as cell invasion \cite{el2019revisiting, maini2004travelling, maini2004traveling}, where well-defined fronts are often observed.

To obtain travelling wave solutiuons with a well-defined front, two main modifications of the Fisher-KPP model have been proposed. The first modification involves incorporating nonlinear degenerate diffusion, whereby the concentration flux is generalized to $ -D(u) \partial_x u$, with $D(0)=0$. One common choice of degenerate diffusivity is $D(u)\equiv u$, which leads to the \textit{Porous-Fisher} model \cite{simpson2006looking, aronson1980density, harris2004fisher, gilding2005fisher, garduno1994approximation, witelski1995shocks, witelski1995merging, sanchezgarduno1995traveling, simpson2011models}:
\begin{equation}
\partial_t u(x,t)= u(x,t)\left[1-u(x,t)\right] +  \partial_{x}\left[u(x,t)\partial_{x}u(x,t)\right], \qquad -\infty < x < \infty. \label{eq:PFSi}
\end{equation}
The Porous-Fisher model supports travelling wave solutions that move with speed $c \ge c_{\textrm{min}}$, where $c_{\textrm{min}} = 1/\sqrt{2}$ \cite{murray, gilding2005fisher, garduno1994approximation, sherratt1996nonsharp}. Phase plane analysis shows that travelling wave solutions of the Porous-Fisher model are sharp-fronted, with compact support, when $c=1/\sqrt{2}$ \cite{murray, garduno1994approximation, sherratt1996nonsharp}.  In contrast, travelling wave solutions of the Porous-Fisher model with $c > 1/\sqrt{2}$ are smooth and do not have compact support \cite{murray, sherratt1996nonsharp}.  Interestingly, travelling wave solutions of the Porous-Fisher model with $c < 1/\sqrt{2}$ have never been reported. 

The second modification of the Fisher-KPP model that leads to a well-defined front is to maintain the use of linear diffusion, but to incorporate a moving boundary condition so that we consider the Fisher-KPP model on $-\infty < x < L(t)$ \cite{du2010spreading, bao2018free, el2019revisiting}. This second modification of the Fisher-KPP model has been called the \textit{Fisher-Stefan} model \cite{du2010spreading, el2019revisiting}, which incorporates a Stefan-like condition at the moving boundary, $L(t)$, to relate the concentration flux, $-\partial_{x}u(x,t)$, with the speed of the moving boundary. One of the limitations of the Fisher-Stefan model is that this model allows populations to become extinct \cite{du2010spreading, el2019revisiting}, since there is an outward flux of at the leading edge, $x=L(t)$.  While the Fisher-Stefan model has the advantage that it can lead to sharp-fronted travelling waves, the physical or biological explanation of the outward flux at $x=L(t)$ is not obvious.

Both the Porous-Fisher and the Fisher-Stefan models have the advantage that they lead to travelling wave solutions with a well-defined front (Figure \ref{fig:schematic}b,c). 
While these two modifications of the Fisher-KPP model have been considered previously, the extension of combining nonlinear degenerate diffusion with a moving boundary condition has yet to be considered. We refer to this combination of the Porous-Fisher model with a moving boundary condition as the \textit{Porous-Fisher-Stefan} model, which we define as the Porous-Fisher equation \eqref{eq:PFSi} on $x\in(-\infty,L(t)]$ with a Stefan-like condition stating that the speed of the moving front, $\mathrm{d}L(t)/\mathrm{d}t$, is proportional to the nonlinear concentration flux, $-u(x,t) \partial_x u(x,t)$, with constant of proportionality $\kappa > 0$. Unlike the Fisher-Stefan model \cite{du2010spreading, el2019revisiting}, the incorporation of nonlinear degenerate diffusion in the Porous-Fisher-Stefan model prevents $u(x,t)$ from being driven to extinction. Interestingly, preliminary numerical solutions of the Porous-Fisher-Stefan model suggest that travelling wave solutions exist with $0\le c < 1/\sqrt{2}$ and that these sharp-fronted travelling waves have infinite slope at $x=L(t)$ (Figure \ref{fig:schematic}d). Neither of these properties have been reported or analyzed previously.

In this work, we examine sharp-fronted travelling waves that arise in the Porous-Fisher-Stefan model. By transforming this model into travelling wave coordinates, we examine the phase plane for various regimes of the wave speed $c$. Using asymptotic analysis, we obtain approximations of the resulting phase plane trajectories when $c\ll1$ and when $c$ is close to the critical wave speed $1/\sqrt{2}$. 
Additionally, we determine the relationship between the travelling wave speed $c$ and the \textcolor{black}{Stefan parameter} $\kappa$. 
In doing so, we determine an approximate form of the travelling wave front that matches numerically-computed travelling wave solutions with high accuracy.

\section{Travelling waves in the Porous-Fisher-Stefan model}
We consider the non-dimensional Porous-Fisher model, describing the concentration $u(x,t)\in[0,1]$ with a Stefan-like condition at the moving boundary $x=L(t)$:
\begin{align}
\partial_t u(x,t)= u(x,t)\left[1-u(x,t)\right]& +  \partial_{x}\left[u(x,t)\partial_{x}u(x,t)\right], \qquad -\infty < x< L(t),\label{eq:PFS}
\\
\lim_{x\to-\infty} &u(x,t)=1,
\qquad
u(L(t),t)=0,
\\
\frac{\mathrm{d}L(t)}{\mathrm{d}t} = -\kappa &\left. u(x,t)\partial_x u(x,t)\right|_{x\to L(t)^-}, \qquad
L(0)=L_0.\label{eq:PFSa}
\end{align}
The Stefan-like condition relates the speed of the moving front, $\mathrm{d}L(t)/\mathrm{d}t$, to the concentration flux, $- u(x,t) \partial_x u(x,t)$, via the constant $\kappa>0$.
Preliminary numerical solutions of \eqref{eq:PFS}--\eqref{eq:PFSa} suggest that travelling waves exist and that these waves move with speed $0\le c < 1/\sqrt{2}$ with infinite gradient at $x=L(t)$. Consequently, we are motivated to examine travelling wave solutions of the Porous-Fisher-Stefan model and to determine the relationship between $c$ and $\kappa$. To do this, we define $\phi(x,t) = [u(x,t)]^2$ to obtain a corresponding PDE with a linear Stefan-like condition at $x=L(t)$:
\begin{align}
\partial_t \phi(x,t) = 2\phi(x,t)\left[1-\sqrt{\phi(x,t)}\right]& +  \sqrt{\phi(x,t)} \, \partial_{xx} \phi(x,t), \qquad -\infty < x< L(t),\label{eq:PFS2}
\\
\lim_{x\to-\infty} &\phi(x,t)=1,
\qquad 
\phi(L(t),t)=0,
\\
\frac{\mathrm{d}L(t)}{\mathrm{d}t} = -&\frac{\kappa}{2} \left. \partial_x \phi(x,t)\right|_{x\to L(t)^-}, \qquad
L(0)=L_0.\label{eq:PFS2a}
\end{align}
To study travelling wave solutions of   \eqref{eq:PFS2}--\eqref{eq:PFS2a}, we transform the system into travelling wave coordinates via $z = x-L_0-ct,$ where $z\in (-\infty,0]$. Noting that when $x=L(t)$, we have $L(t)=L_0+ct$, and hence,
$\mathrm{d}L(t)/\mathrm{d}t = c$. This change of coordinates gives
\begin{align}
\sqrt{\phi(z)}\,\phi''(z) + &c\phi'(z) + 2\phi(z)\left(1-\sqrt{\phi(z)}\right)=0, \qquad -\infty<z<0, \label{eq:TWode}
\\
\lim_{z\to-\infty} \phi(z)=1, \quad &\phi(0)=0, \quad  \lim_{z\to0^-}\phi'(z) = -\frac{2c}{\kappa}, \label{eq:TWbc}
\end{align}
where $\displaystyle ' = \mathrm{d}/\mathrm{d}z$. It is convenient to write \eqref{eq:TWode} as a system of two first order differential equations:
\begin{align}
\phi'(z) &= \psi(z), \label{eq:sys1}
\\
\psi'(z) &=-\frac{c\psi(z)}{\sqrt{\phi(z)}}-2\sqrt{\phi(z)}\left(1-\sqrt{\phi(z)}\right). \label{eq:sys2}
\end{align}
A general explicit solution of \eqref{eq:sys1}--\eqref{eq:sys2} is not obvious, so we seek to study solutions of this system using the $(\phi(z), \psi(z))$ phase plane. There are only two fixed points of \eqref{eq:sys1}--\eqref{eq:sys2}: $(0,0)$ and $(1,0)$. Typically, with phase plane analysis of travelling wave solutions, we 
explore the possibility of a heteroclinic orbit between these fixed points by considering the local behaviour of the linearized system near the fixed points \cite{murray, sanchezgarduno1995traveling}. For the Porous-Fisher model, whose phase plane is identical to \eqref{eq:sys1}--\eqref{eq:sys2}, such heteroclinic orbits can only occur for $c\ge 1/\sqrt{2}$  \cite{murray, sherratt1996nonsharp}. However, for the Porous-Fisher-Stefan model, we have the addition of the Stefan-like condition in \eqref{eq:TWbc}. Therefore, travelling wave solutions of the Porous-Fisher-Stefan model do not correspond to a heteroclinc orbit between $(0,0)$ and $(1,0)$, but rather a trajectory between $(1,0)$ and another point that is determined by the moving boundary condition in \eqref{eq:TWbc}. This particular trajectory, $\psi(\phi;c)$, corresponds to the travelling wave solution for a given $c$ with $\displaystyle \lim_{z\to-\infty} \phi(z)=1$, as well as determining $\kappa$ via \eqref{eq:TWbc}: $\kappa = -2c/\psi(0;c)$.

 To determine $\psi(\phi;c)$, we divide \eqref{eq:sys2} by \eqref{eq:sys1} and obtain
\begin{equation}
-\frac{\mathrm{d}\psi(\phi)}{\mathrm{d}\phi} = \frac{c}{\sqrt{\phi}}+\frac{2\sqrt{\phi}\left(1-\sqrt{\phi}\right)}{\psi(\phi)}, \qquad \psi(1)=0, \qquad \lim_{\phi\to0^+} \psi(\phi)=-\frac{2c}{\kappa}. \label{eq:dvdu}
\end{equation}
As previously mentioned, physical travelling wave solutions in the Porous-Fisher model can only occur when $c \ge 1/\sqrt{2}$ \cite{murray, sherratt1996nonsharp}, and numerical simulations of the Porous-Fisher-Stefan model indicate that all travelling waves have $c < 1/\sqrt{2}$. As a result, we examine \eqref{eq:dvdu} in two limiting regimes: $c \to 0^+$ and $c \to 1/\sqrt{2}\,^-$.

\subsection{Travelling wave solutions for $c\ll 1$}
We first consider the solution of \eqref{eq:dvdu} in the limit where $0\le c\ll 1$ by expanding $\psi(\phi)$ as a regular perturbation expansion in $c$, i.e. 
$\psi(\phi) = V_0(\phi) + c V_1(\phi) + \mathcal{O}(c^2)$.
Substituting this expansion into \eqref{eq:dvdu} provides 
\begin{align}
&\mathcal{O}(1): \qquad -V_0(\phi)\frac{\mathrm{d}V_0(\phi)}{\mathrm{d}\phi} = 2\sqrt{\phi}\left(1-\sqrt{\phi}\right),  &V_0(1)=0, \label{eq:O1}
\\
&\mathcal{O}(c): \qquad -V_1(\phi)\frac{\mathrm{d}V_0(\phi)}{\mathrm{d}\phi}-V_0(\phi)\frac{\mathrm{d}V_1(\phi)}{\mathrm{d}\phi} =\frac{V_0(\phi)}{\sqrt{\phi}},  &V_1(1)=0. \label{eq:Oc}
\end{align}
The solution of \eqref{eq:O1}--\eqref{eq:Oc} are
\begin{equation}
V_0(\phi) = -\sqrt{\frac{2}{3}-\frac{8}{3}\phi^{3/2}+2\phi^2} 
~~\text{ and }~~
 V_1(\phi) = -\frac{1}{V_0(\phi)} \int_\phi^1 \sqrt{\frac{2}{3s}-\frac{8}{3}\sqrt{s}+2s} \, \mathrm{d}s.
\end{equation}
Evaluating these expressions at $\phi=0$ gives a two-term approximation for the wave speed $c$ as a function of $\kappa$, provided that $c\ll 1$:
\begin{equation}
\kappa \sim \frac{54\sqrt{3}c}{27\sqrt{2}-\alpha c} \iff c\sim \frac{27\sqrt{2}\kappa}{54\sqrt{3}+\alpha\kappa}, \label{eq:ck0}
\end{equation}
where $\alpha = 36\sqrt{2}-6\sqrt{3}+24\log[(\sqrt{3}-1)/(3\sqrt{2}-4)] \approx 67.02$.
We note that since $\psi = 2u u'$ remains $\mathcal{O}(1)$ as $\phi \to 0^+$, this implies that $u' = \mathcal{O}(u^{-1})$ as $u \to 0^+$, confirming that we are examining a new class of travelling wave solutions with infinite slope at the moving boundary.

\subsection{Travelling wave solutions for $c\to 1/\sqrt{2}\,^-$}

From \cite{murray, garduno1994approximation, sherratt1996nonsharp}, we know that when $c=1/\sqrt{2}$, \eqref{eq:dvdu} has the solution $u' = (u-1)/\sqrt{2}$, implying that $\psi=2uu' \to 0$ as $u\to0^+$. The Stefan-like condition in \eqref{eq:dvdu} requires that $-2c/\kappa$ must also equal zero. Since $c\ne 0$, this implies that $\kappa \to \infty$ as $c \to 1/\sqrt{2}\,^-$. To determine the leading-order behaviour of $\kappa$ in this limit, we let $c=1/\sqrt{2}-\varepsilon$, with $0\le \varepsilon \ll 1$, and \eqref{eq:dvdu} becomes
\begin{equation}
-\frac{\mathrm{d}\psi(\phi)}{\mathrm{d}\phi} = \frac{1}{\sqrt{2\phi}}-\frac{\varepsilon}{\sqrt{\phi}}+\frac{2\sqrt{\phi}\left(1-\sqrt{\phi}\right)}{\psi(\phi)}, \quad 
\psi(1)=0, \quad \lim_{\phi\to0^+} \psi(\phi)=-\frac{2}{\kappa}\left(\frac{1}{\sqrt{2}}-\varepsilon\right). \label{eq:psiODE}
\end{equation}
We perform a regular perturbation expansion in $\varepsilon$, i.e.
$\psi = \Psi_0(\phi) + \varepsilon\Psi_1(\phi)+ \mathcal{O}(\varepsilon^2)$,
which gives
\begin{align}
&\mathcal{O}(1): \qquad -\frac{\mathrm{d}\Psi_0(\phi)}{\mathrm{d}\phi} = \frac{1}{\sqrt{2\phi}}+\frac{2\sqrt{\phi}\left(1-\sqrt{\phi}\right)}{\Psi_0(\phi)}, \qquad &\Psi_0(1)=0, \label{eq:psi0}
\\
&\mathcal{O}(\varepsilon): \qquad -\Psi_1(\phi)\frac{\mathrm{d}\Psi_0(\phi)}{\mathrm{d}\phi} -\Psi_0(\phi)\frac{\mathrm{d}\Psi_1(\phi)}{\mathrm{d}\phi}= \frac{\Psi_1(\phi)}{\sqrt{2\phi}}-\frac{\Psi_0(\phi)}{\sqrt{\phi}}, \qquad &\Psi_1(1)=0. \label{eq:psi1}
\end{align}
Noting that $u_0'=(u_0-1)/\sqrt{2}$ is the leading-order solution in the original variables, the solutions of \eqref{eq:psi0}--\eqref{eq:psi1} are
\begin{equation}
\Psi_0(\phi) = 2u_0u'_0= -\sqrt{2\phi}\left(1-\sqrt{\phi}\right)
~~\text{ and }~~
\Psi_1(\phi)=-\frac{2\left(1-\sqrt{\phi}\right)}{3}.
\end{equation}
Evaluating $\Psi_0, \Psi_1$ at $\phi=0$ retrieves the Stefan-like condition
 in \eqref{eq:psiODE}; hence, for $c = 1/\sqrt{2}-\varepsilon$, corresponding to $\kappa\gg 1$, we have 
 \begin{equation}
 \kappa\sim \frac{3\sqrt{2}c}{1-\sqrt{2}c} \iff c \sim \frac{\kappa}{\sqrt{2}\,(\kappa+3)}.\label{eq:ckinf}
 \end{equation}
 In the limit where $\kappa \gg 1$, we can also approximate the travelling wave front $\phi(z)$. To do this, we note that
 \begin{equation}
 \frac{\mathrm{d} \phi}{\mathrm{d}z} \sim \Psi_0(\phi)+\varepsilon \Psi_1(\phi) = \sqrt{2}\left(\sqrt{\phi}-1\right)\left(\sqrt{\phi}+\frac{\varepsilon \sqrt{2}}{3}\right), \qquad \phi(0)=0,
 \end{equation}
with solution
  \begin{equation}
 z(\phi) = \frac{3\sqrt{2}}{3+\varepsilon \sqrt{2}}\left[\log\left(1-\sqrt{\phi}\right)+\frac{\varepsilon \sqrt{2}}{3}\log\left(1+\frac{3}{\varepsilon }\sqrt{\frac{\phi}{2}}\right)\right].
 \end{equation}
Equivalently, the original travelling wave $u(z)=\sqrt{\phi(z)}$ can be written implicitly, using \eqref{eq:ckinf}, as
 \begin{equation}
 z(u) \sim \frac{(\kappa+3)\sqrt{2}}{\kappa+4}\left[\log\left(1-u\right)+\frac{\log\left(1+(\kappa+3)u\right)}{\kappa+3}\right]. \label{eq:ZuKap}
 \end{equation}
 Thus, we can approximate the sharp moving front $u(z)$, along with its corresponding wave speed $c$, in the regime where $\kappa$ is large. Furthermore, we note that in the limit where $\kappa \to \infty$, we have $z(u) \to \sqrt{2}\log(1-u)$, implying that $u(z) = 1-\exp\left(z/\sqrt{2}\right)$. This result agrees with the sharp-fronted travelling wave determined in \cite{murray, garduno1994approximation, sherratt1996nonsharp}.

\subsection{Comparison of travelling wave solutions}
 
\begin{figure}
\begin{center}
\includegraphics[width= \textwidth]{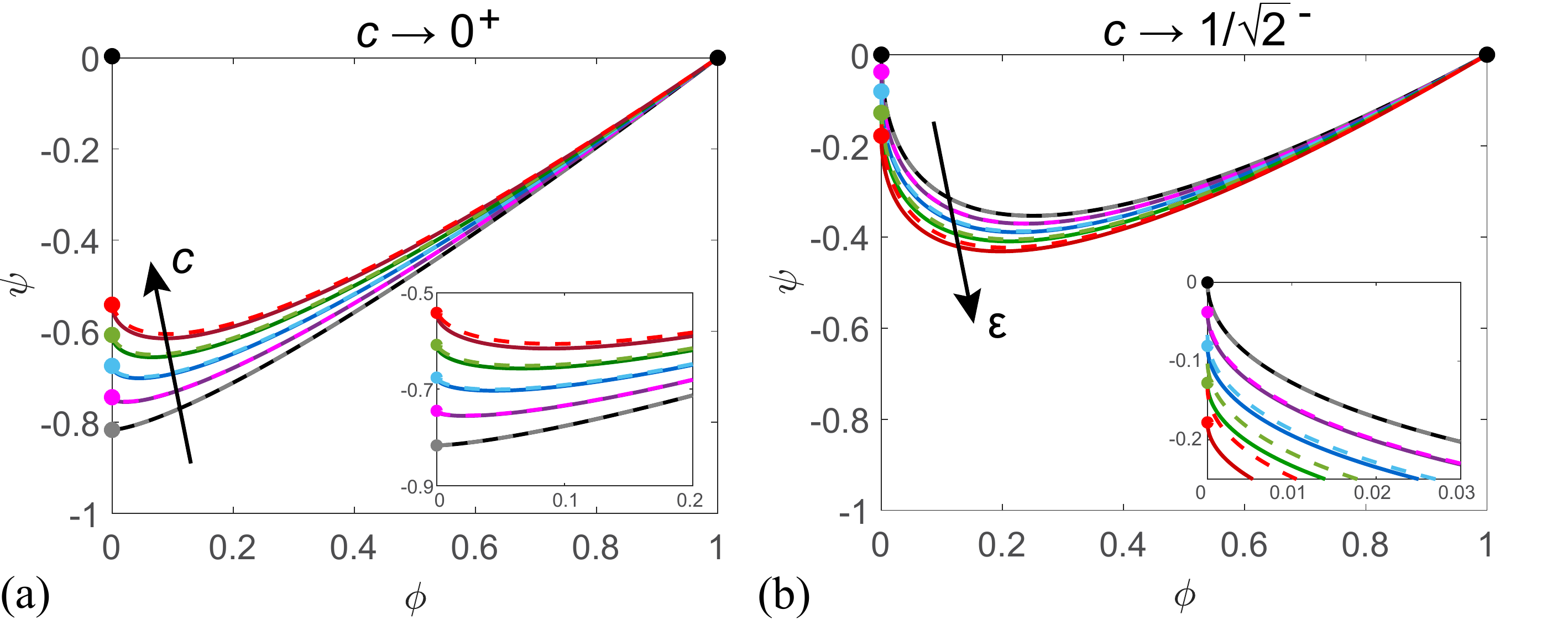}
\end{center}
\vspace{-0.2in} 
 \caption{Comparison of the trajectory $\psi(\phi;c)$ in the $(\phi,\psi)$-phase plane (solid curves), which solves \eqref{eq:dvdu} for a given $c$, with its two-term asymptotic approximations (dashed curves). The fixed points $(\phi,\psi)=(0,0), (1,0)$ are shown as  black circles, while the Stefan-like condition at $\phi=0$ is shown as coloured circles. (a) The two-term asymptotic approximation when $c\ll1$ is $V_0+cV_1$, where $c=0$ (black/grey), $c=0.05$ (pink), $c=0.1$ (blue), $c=0.15$ (green), and $c=0.2$ (red). \textcolor{black}{The black arrow points in the direction of increasing $c$.} (b) The two-term asymptotic approximation when $c= 1/\sqrt{2}-\varepsilon$ is $\Psi_0+\varepsilon\Psi_1$, where $\varepsilon=0$ (black/grey), $\varepsilon=0.05$ (pink), $\varepsilon=0.1$ (blue), $\varepsilon=0.15$ (green), and $\varepsilon=0.2$ (red). The insets in (a) and (b) show the solutions near $\phi=0$. \textcolor{black}{The black arrow points in the direction of increasing $\varepsilon$.}\label{fig:pp}}
\end{figure}

To validate our asymptotic approximations, we firstly examine the trajectory $\psi(\phi;c)$, which solves \eqref{eq:dvdu} for a given $c$, in the $(\phi,\psi)$-phase plane. To determine $\psi(\phi;c)$, we solve \eqref{eq:dvdu} numerically using \href{https://au.mathworks.com/help/matlab/ref/ode45.html}{\texttt{ode45}} in MATLAB. In Figure \ref{fig:pp}a, the two-term approximation $V_0(\phi)+cV_1(\phi)$, valid when $c\ll1$, agrees very well with $\psi(\phi;c)$ up to $c=0.2$. Furthermore, in Figure \ref{fig:pp}b, we have good agreement between $\psi(\phi;c)$ and the two-term approximation $\Psi_0(\phi)+\varepsilon\Psi_1(\phi)$, in the limit when $c=1/\sqrt{2}-\varepsilon$ up to $\varepsilon=0.2$.

\begin{figure}
\begin{center}
\includegraphics[width= .9\textwidth]{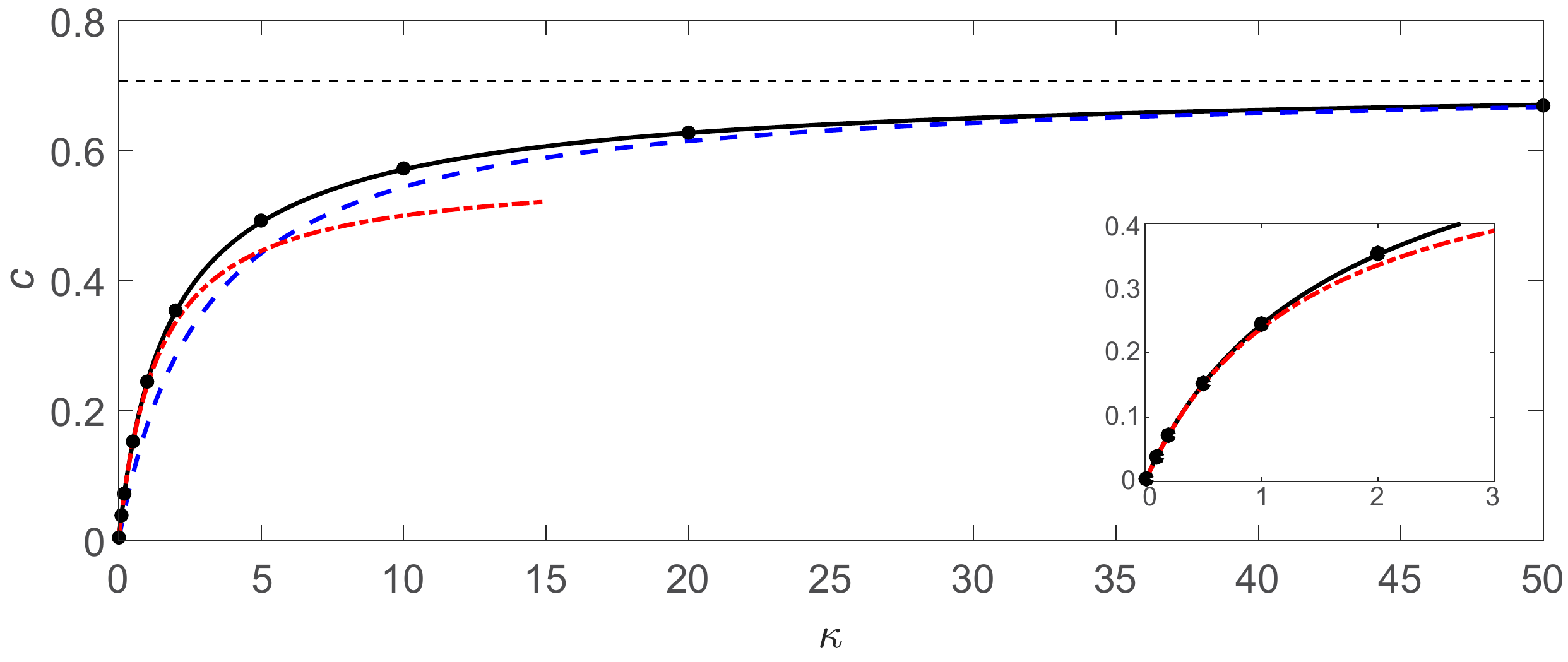}
\end{center}
\vspace{-0.2in} 
 \caption{Comparison of the $c(\kappa)$ curves: numerically computed from \eqref{eq:PFS2}--\eqref{eq:PFS2a} (black circles), numerically computed from \eqref{eq:dvdu} \textcolor{black}{(black solid curve)}, and the two-term approximations for $c\ll1$ (red dot-dash curve) and $\kappa \gg 1$ (blue dashed curve). The critical wave speed $c= 1/\sqrt{2}$ is shown as a black dashed line. \label{fig:stef}}
\end{figure}

Since the asymptotic approximations accurately describe $\psi(\phi;c)$ in the phase plane, we now examine the asymptotic relationship between $c$ and $\kappa$. We can estimate $c(\kappa)$ from the numerically-computed $\psi(\phi;c)$ by noting that $c = -\kappa \,\psi(0;c)/2$. Alternatively, we can use the numerical solutions of \eqref{eq:PFS2}--\eqref{eq:PFS2a} to determine the speed of the moving front $L(t)$, once the solution has settled to the travelling wave solution (see Appendix A). Figure \ref{fig:stef} shows that these two numerical approaches to determine $c(\kappa)$ are indeed equivalent. Furthermore, we see that both asymptotic approximations for $c(\kappa)$, \eqref{eq:ck0} and \eqref{eq:ckinf}, agree with the numerically-determined $c(\kappa)$ relationship. To minimize the error between $c(\kappa)$ and its asymptotic approximations, we recommend that \eqref{eq:ck0} be used for $\kappa<2$ and \eqref{eq:ckinf} be used for $\kappa>20$. For intermediate values of $\kappa$, it appears that a numerical solution is warranted.

\begin{figure}
\begin{center}
\includegraphics[width= \textwidth]{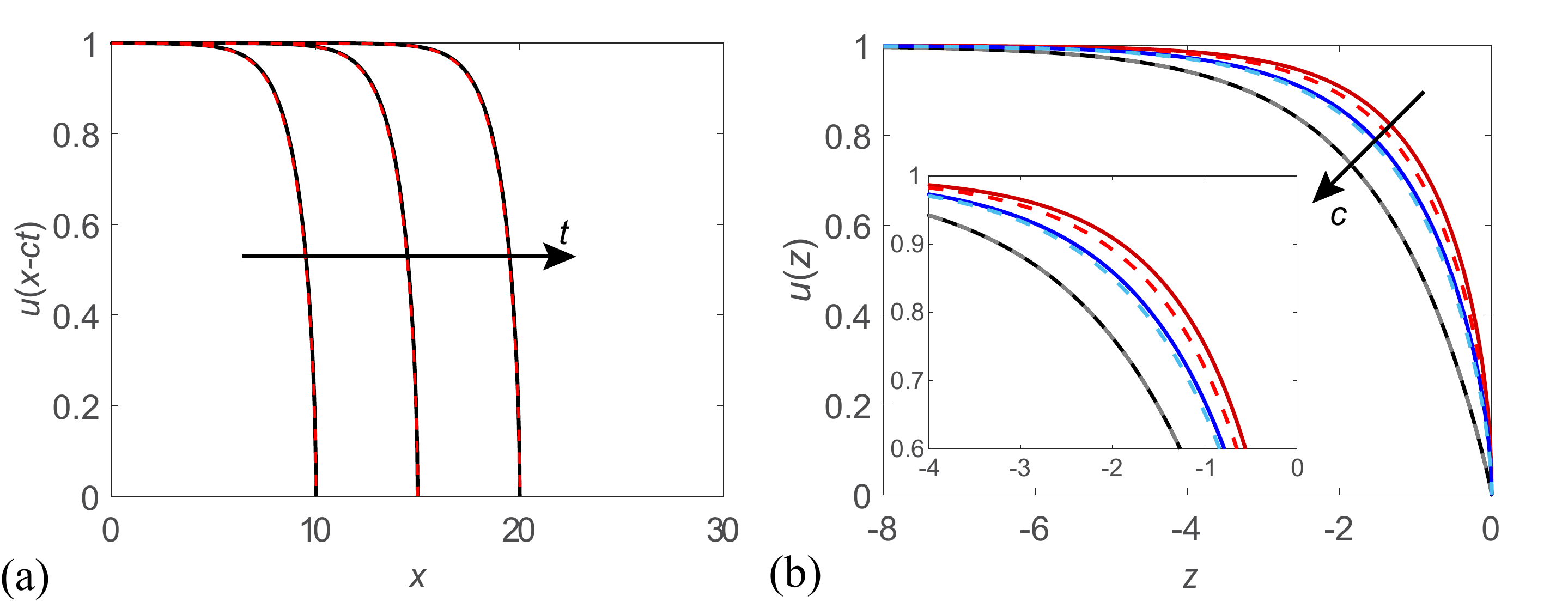}
\end{center}
\vspace{-0.2in} 
 \caption{(a) Comparison of the travelling wave front $u(x-ct)$ for $\kappa=1$. The numerical solution of \eqref{eq:PFS2}--\eqref{eq:PFS2a} is shown in black at equally-spaced times ($t=41.0, 61.5, 82.0$), using \eqref{eq:ZuKap} as an initial condition. The red-dashed front is determined from large-$\kappa$ asymptotics \eqref{eq:ZuKap}.  \textcolor{black}{The black arrow points in the direction of increasing $t$.} (b) Travelling wave fronts $u(z)$ (solid curves), superimposed on the corresponding asymptotic approximations (dashed curves) determined using \eqref{eq:ckinf} and \eqref{eq:ZuKap}, for $c=0.7$ (black/grey), $c=0.4$ (blue), and $c=0.1$ (red). \textcolor{black}{The corresponding values of $\kappa$ are 287.1 (295.5), 2.711 (3.907), and 0.2958 (0.4941), respectively, where numerically-determined values are shown in the text and asymptotic approximations of $\kappa$ using \eqref{eq:ckinf} are shown in parentheses. The black arrow points in the direction of increasing $c$.}\label{fig:fronts}}
\end{figure} 

Finally, we compare the asymptotic approximation of the moving front $u(z)=u(x-ct)$, shown in \eqref{eq:ZuKap}, with the numerical solution of \eqref{eq:PFS2}--\eqref{eq:PFS2a}, after sufficient time has passed such that the travelling wave has formed (Appendix A).
\textcolor{black}{Despite the fact that the asymptotic approximation is only} expected to match when $\kappa$ is large, we see in Figure \ref{fig:fronts}a that the asymptotic approximation of the shape of the moving fronts, shown implicitly in \eqref{eq:ZuKap}, matches the numerically-determined travelling wave solution very well, even when $\kappa=1$. Furthermore, the asymptotic approximation in \eqref{eq:ZuKap} agrees well with the numerically-computed travelling wave fronts for a variety of wave speeds (Figure \ref{fig:fronts}b).

\section{Conclusions}

In this work, we consider new travelling wave solutions for the Porous-Fisher model with a moving boundary. This model has certain features that could be considered to be advantageous over other modifications of the Fisher-KPP model. In summary, the new travelling wave solutions: (i) move with a speed that is slower than the more standard Porous-Fisher model; (ii) never lead to population extinction; (iii) have compact support and a well-defined moving front, and (iv) the travelling wave profiles have an infinite slope at the moving front. Using travelling wave coordinates, we transform the model into a single nonlinear differential equation. The solution of this differential equation, corresponding to a particular trajectory in the associated phase plane, determines the travelling wave front and can be approximated using asymptotic analysis in two limiting parameter regimes. In both cases, we obtain a good approximation of this trajectory, which can also be used to relate the speed of the moving front to $\kappa$, a constant appearing in the Stefan-like condition. Finally, we determine a highly-accurate approximate form of the sharp-fronted travelling wave with infinite slope at the moving boundary, corresponding to solutions with wave speeds $0\le c<1/\sqrt{2}$.

Further extensions of this Porous-Fisher-Stefan model can also be made. For instance, the asymptotic analysis performed in this work can be extended to include higher-order terms.
Another possible extension could be to consider generalizing the nonlinear degenerate diffusivity function to $D(u)=u^n$, for some constant $n>0$ \cite{simpson2006looking, aronson1980density, gilding2005fisher, wang2008integrability, mccue2019hole}.  We leave both extensions for future consideration.

\subsection*{Acknowledgements}
This work is supported by the Australian Research Council (DP170100474). 

\newpage
\appendix
\section{Numerical Solution for the Porous-Fisher-Stefan model}

A MATLAB implementation of the numerical solution of \eqref{eq:PFS2}--\eqref{eq:PFS2a}, described below, can be found at \url{https://github.com/nfadai/Fadai_TW2019}.

To numerically compute solutions of \eqref{eq:PFS2}--\eqref{eq:PFS2a}, we first approximate the semi-infinite domain $(-\infty, L(t)]$ as the finite domain  $[0, L(t)]$, provided that $L_0>0$. Additionally, we transform \eqref{eq:PFS2}--\eqref{eq:PFS2a} to a fixed-space domain by setting 
$\xi = x/L(t).$
Consequently,\eqref{eq:PFS2}--\eqref{eq:PFS2a} becomes

\begin{align}
&\partial_t \phi(\xi,t) = \frac{\xi}{L(t)} \, \frac{\mathrm{d}L(t)} {\mathrm{d}t}\,\partial_{\xi} \phi(\xi,t) + 2\phi(\xi,t)\left[1-\sqrt{\phi(\xi,t)}\right] + \frac{\sqrt{\phi(\xi,t)} }{[L(t)]^2} \, \partial_{\xi \xi} \phi(\xi,t), \label{eq:PFSq}
\\
&\left.\partial_{\xi}\phi(\xi,t)\right|_{\xi=0}=0,
\qquad 
\phi(1,t)=0,
\qquad
0<\xi<1,
\\
&L(t)\frac{\mathrm{d}L(t)}{\mathrm{d}t} = -\frac{\kappa}{2} \left. \partial_{\xi} \phi(\xi,t)\right|_{\xi=1}, \qquad
L(0)=L_0.\label{eq:PFSq2}
\end{align}
To compute the numerical solution of \eqref{eq:PFSq}--\eqref{eq:PFSq2}, we must specify values for $L_0$, $\kappa$ and $\phi(\xi,0)$. We obtain numerical solutions of \eqref{eq:PFSq} on a uniformly-spaced mesh of $\xi\in[0,1]$, i.e. $\xi_i = i \Delta \xi$, $i=0, \dots, N$, where $\Delta \xi = 1/N$. We denote $\phi(\xi_i,t_j)=\phi_i^n$ and $L(t_j)=L_j$ for convenience, where $n \ge 1$ is the $n$th Picard iteration estimate at time $t_j$. Therefore, to determine $\phi_i$, we use

\begin{align}
\frac{\phi_i^{n}-\phi_i^p}{\Delta t} = &\frac{\xi_i (L_j-L_{j-1})(\phi_{i+1}^n-\phi_{i-1}^n)}{2 L_j \Delta t \Delta \xi} + 2\phi_{i}^n\left[1-\sqrt{\phi_{i}^{n-1}}\right] 
\label{eq:Num}
\\
&+ \frac{(\phi_{i+1}^n-2\phi_{i}^n+\phi_{i-1}^n)\sqrt{\phi_{i}^{n-1}}}{(L_j)^2 (\Delta \xi)^2} , \notag
\\
&\phi_{0}^n=\phi_{1}^n,
\qquad 
\phi_N^n=0.\label{eq:Num2}
\end{align}
Here, $\phi_i^p$ is the solution of $\phi_i$ at the previous timestep, $t_{j-1}$, and $\Delta t$ is the timestep.
We identify the system \eqref{eq:Num}--\eqref{eq:Num2} as a tridiagonal matrix in $\phi_i^n$ at time $t_j$, which we can solve efficiently using the Thomas algorithm. This solution is stored as $\pmb{\Phi}_n$; if $\displaystyle \max \left| \pmb{\Phi}_n - \pmb{\Phi}_{n-1}\right|<\delta$, where $\delta$ is some user-specified tolerance, then the Picard loop terminates and we proceed to updating the moving boundary for the next timestep. Otherwise, $n=n+1$, the solution $\pmb{\Phi}_n$ is stored as $\phi_i^{n-1}$, and the Picard iteration loop is performed again.


From the fixed-boundary PDE, the Stefan-like condition at $\xi=1$ is \begin{equation}
L(t)\frac{\mathrm{d}L(t)}{\mathrm{d}t} = -\frac{\kappa}{2} \left. \partial_{\xi} \phi(\xi,t)\right|_{\xi=1}, \qquad t\in [t_{j},t_{j+1}], \qquad
L(t_{j})=L_{j}. \label{eq:Stef}
\end{equation}
We approximate $\phi(\xi,t)$ as $\phi(\xi,t_j)$ during the small interval $t \in [t_j, t_{j+1}]$, allowing us to explicitly solve \eqref{eq:Stef} to give a closed form approximation for $L(t)$:
\begin{equation}
L(t) = \sqrt{(L_j)^2-\kappa \left. \partial_{\xi} \phi(\xi,t_j)\right|_{\xi=1}(t-t_j)}, \qquad t_j \le t \le t_{j+1}.
\end{equation}
Therefore, evaluating this expression at $t=t_{j+1}$ and using a second-order finite difference approximation of $ \displaystyle \left. \partial_{\xi} \phi(\xi,t_j)\right|_{\xi=1}$, we obtain the approximation
\begin{equation}
L_{j+1} = \sqrt{(L_j)^2-\frac{\kappa \Delta t (3\phi_N^n-4\phi_{N-1}^n+\phi_{N-2}^n)}{2\Delta \xi}}.
\end{equation}
With these updated values of $L_{j+1}$ and $\pmb{\Phi}_n = \phi_i^p$, we update $t=t+\Delta t$ and $j=j+1$; we then repeat the computation to integrate through the next time increment. The algorithm terminates when $t+\Delta t > t_f$, where $t_f$ is the user-specified final time.
Once sufficient time has passed that the solution settles towards a travelling wave, we expect that $\mathrm{d}L(t)/\mathrm{d}t = c$, so we fit a straight line to our numerical estimate of $L(t)$ and use the slope of that line to provide an estimate of $c$.

\end{document}